\documentclass[useAMS,usenatbib]{mn2e}
\usepackage{graphicx}
\usepackage{url}

\def\aap{\ {A\&A}\ }

\def\aj{\ {AJ}\ }

\def\apj{\ {ApJ}\ }
\def\apjl{\ {ApJL}\ }
\def\apjs{\ {ApJS}\ }

\def\araa{\ {ARA\&A}\ }

\def\mnras{\ {MNRAS}\ }
\def\nat{\ {Nat}\ }

\def\pasp{\ {PASP}\ }

\def\aap{\ {A\&A}\ }

\def\aj{\ {AJ}\ }

\def\apj{\ {ApJ}\ }
\def\apjl{\ {ApJL}\ }
\def\apjs{\ {ApJS}\ }

\def\araa{\ {ARA\&A}\ }

\def\mnras{\ {MNRAS}\ }
\def\nat{\ {Nat}\ }

\def\na{\ {New Astron.}\ }

\def\pasp{\ {PASP}\ }

\newcommand{\MSun}{\mbox{${\rm M}_\odot$}}
\newcommand{\Msun}{\mbox{${\rm M}_\odot$}}

\newcommand{\LSun}{\mbox{${\rm L}_\odot$}}

\newcommand{\RSun}{\mbox{${\rm R}_\odot$}}

\def\lteq{\ {\raise-.5ex\hbox{$\buildrel<\over-$}}\ }
\def\apgt{\ {\raise-.5ex\hbox{$\buildrel>\over\sim$}}\ }
\def\aplt{\ {\raise-.5ex\hbox{$\buildrel<\over\sim$}}\ }
\def\lt{\ {\raise-.5ex\hbox{$\buildrel>$}}\ }
\def\gt{\ {\raise-.5ex\hbox{$\buildrel<$}}\ }
\def\eqgt{\ {\raise-.5ex\hbox{$\buildrel>\over-$}}\ }
\def\eqlt{\ {\raise-.5ex\hbox{$\buildrel<\over-$}}\ }

\newfont{\Giga}{cmssbx10 scaled 5200}
\newfont{\giga}{cmssbx10 scaled 4300}
\newfont{\Mega}{cmssbx10 scaled 3200}
\newfont{\mega}{cmssbx10 scaled 2500}
\newfont{\Kilo}{cmssbx10 scaled 2000}
\newfont{\kilo}{cmssbx10 scaled 1600}
\newfont{\Deca}{cmssbx10 scaled 1450}
\newfont{\deca}{cmssbx10 scaled 1200}
\newfont{\Dezi}{cmssbx10 scaled 1100}
\newfont{\dezi}{cmssbx10 scaled 1050}
\newfont{\iGiga}{cmssi10 scaled 6200}
\newfont{\igiga}{cmssi10 scaled 4300}
\newfont{\iMega}{cmssi10 scaled 3200}
\newfont{\imega}{cmssi10 scaled 2500}
\newfont{\iKilo}{cmssi10 scaled 2000}
\newfont{\ikilo}{cmssi10 scaled 1500}
\newfont{\mathGiga}{cmsy10 scaled 6200}
\newfont{\mathgiga}{cmsy10 scaled 4300}
\newfont{\mathMega}{cmsy10 scaled 3200}
\newfont{\mathmega}{cmsy10 scaled 2500}
\newfont{\mathKilo}{cmsy10 scaled 2000}
\newfont{\mathkilo}{cmsy10 scaled 1500}
\newfont{\mathDeca}{cmsy10 scaled 1450}
\newfont{\mathdeca}{cmsy10 scaled 1200}

\def\apgt{\ {\raise-.5ex\hbox{$\buildrel>\over\sim$}}\ }
\def\aplt{\ {\raise-.5ex\hbox{$\buildrel<\over\sim$}}\ }
\def\lteq{\ {\raise-.5ex\hbox{$\buildrel<\over-$}}\ }

\title[]{Was the nineteenth century giant eruption of Eta Carinae a
  merger event in a triple system?}

\author[]{S.F. Portegies Zwart$^{1}$ and E.P.J. van den Heuvel$^{2,3}$\\
$^{1}$Leiden Observatory, Leiden University, PO Box 9513, 2300 RA, Leiden, 
The Netherlands \\
$^{2}$ Astronomical Institute ‘Anton Pannekoek’, University of Amsterdam, 
P.O. Box 94249, 1090 GE, Amsterdam, The Netherlands \\
$^{3}$Kavli Institute for Theoretical Physics, UCSB, Santa Barbara, CA 93106-4030}

\begin{document}
%\baselineskip=1.5\baselineskip

\date{}
%\pagerange{ -- } \pubyear{2011} 
\maketitle

\begin{abstract} %%<300 words

  We discuss the events that led to the giant eruption of Eta Carinae,
  and find that the mid-nineteenth century (in 1838-1843) giant
  mass-loss outburst has the characteristics of being produced by the
  merger event of a massive close binary, triggered by the
  gravitational interaction with a massive third companion star, which
  is the current binary companion in the Eta Carinae system.  We come
  to this conclusion by a combination of theoretical arguments
  supported by computer simulations using the Astrophysical
  Multipurpose Software Environment. According to this model the $\sim
  90$\,\MSun\, present primary star of the highly eccentric Eta
  Carinae binary system is the product of this merger, and its $\sim
  30$\,\MSun\, companion originally was the third star in the system.

  In our model the Homunculus nebula was produced by an extremely
  enhanced stellar wind, energized by tidal energy dissipation prior
  to the merger, which enormously boosted the radiation-driven wind
  mass-loss.  The current orbital plane is then aligned with the
  equatorial plane of the Homunculus, and the symmetric lobes are
  roughly aligned with the argument of periastron of the current Eta
  Carina binary. The merger itself then occurred in 1838, which
  resulted in a massive asymmetric outflow in the equatorial plane of
  the Homunculus. The 1843 outburst can in our model be attributed to
  the subsequent encounter when the companion star (once the outer
  most star in the triple system) plunges through the bloated envelope
  of the merger product, once when it passed periastron again.  We
  predict that the system has an excess space velocity of order
  50\,km/s in the equatorial plane of the Homunculus. Our triple model
  gives a viable explanation for the high runaway velocities typically
  observed in LBVs \citep{2015MNRAS.447..598S}.

\end{abstract}


%\citep{2014ApJ...796..121J} Luminous Blue Variables and Super-luminous Supernovae from Binary Mergers\\
%\citep{1989ASSL..157..185G} CE evoelution for LBVs\\

\section{Introduction}

Today, Eta Carinae is a binary system, composed of a $\sim
90\,\MSun$\, primary star with an estimated age of $\aplt 1$\, Myr
\citep{1997ARA&A..35....1D}, although the current copious mass loss
rate of $1.6 \times 10^{-3}$\,\MSun/yr \citep{2003A&A...410L..37V},
might indicate an older age, possibly on the verge of turning into a
Wolf-Rayet star \citep{2006ApJ...645L..45S}.  The primary is orbited
by a 30\,\MSun\, presumably main-sequence star in a 15.46\,AU orbit
with an eccentricity of about 0.9 \citep{1997NewA....2..107D}.

In the mid 19-th century Eta Carinae experienced a giant eruption.
Evidence for this is still visible in its surrounding nebula
\citep{1999ASPC..179..216H}.  During this event it outshone any star
in the sky except Sirius \citep{1952PASP...64..185D}.  There is no
satisfactory scenario that explains the copious mass loss during the
19th century giant eruption, and there is no ready explanation for the
at least 2 magnitude bolometric brightening of the object in the
$\apgt 30$ years before the great eruption
\citep{2013MNRAS.429.2366S}. We propose that both are related, and
caused by the eventual merger of a massive close binary, which was
triggered by gravitational interaction with a third star in a wide
orbit.  This third component became Eta Carinae's binary companion,
after the merger event. This model has a number of attractive
characteristics and interesting consequences for other massive triples
to produce objects like Eta Carinae.

Eta Carinae is extraordinary in many respects.  The system is
  often referred to as a 'supernova impostor', and possible candidates
  that may have met a similar fate include 
  SN1954J \cite{2001PASP..113..692S}, 
  V838 Mon \citep{2002A&A...389L..51M},
  SN2009ip \citep{2009CBET.1928....1M}, 
  SN1961v \citep{2012ApJ...758..142K},
  and $\delta$ Scorpius \citep{2013ApJ...766..119M}. 
  The comparison with the outburst in
  SN2009ip \citep{2011MNRAS.415..773S} may be particularly important
  in understanding the diversity in spectral properties of eruptions
  in luminous blue variables in general \citep{2011ApJ...732...32F}.
  The high mass and high velocity of the great eruption of Eta Carinae
  and the outburst in SN2009ip may then have a similar origin,
  possibly resembling that of a common envelope event
  \citep{2013MNRAS.430.1801M,2013ApJ...777L..35T,2014AJ....147...23L}.
  These events are referred to as ``mergerburst'' by
  \cite{2013ApJ...764L...6S}.  Supernova impostor PTF11iqb was
  initially classified as a type SN II-N \citep{2011ATel.3510....1P}
  but with weaker pre-supernova mass loss, as seen in SNe II-L and
  II-P, but was later characterised as an event similar to Eta Carinae
  \citep{2015arXiv150102820S}.  Similarly, the NGC~300 OT2008-1 event
  was suggested to be a scaled down version of the same process that
  drove the great eruption in Eta Carinae
  \citep{2010ApJ...709L..11K}.

Potential future candidates include HD75821, HD167263 and HD135240
\citep{2008MNRAS.389..869E,2010yCat..73890869E}, and the circumstellar
nebular R4 has characteristics of having experienced a similar binary
merger \citep{2000AJ....119.1352P}.

Several of today's parameters of Eta Carinae are well
constrained by the observations, but what initiated the great eruption
and how the system looked like before this event is actively
  debated among astronomers.  In this paper we aim at
  constraining the parameters and the events that led to the giant
  eruption, but we start constraining some of the current
  observables. For example, to avoid violation of the Eddington limit
for Eta Carina's luminosity ($L_{\rm bol} = 10^{6.7}$\,\LSun) during
its great eruption, its mass must be $\apgt 80$\,\MSun
\citep{2012ASSL..384....1H}. The star has a large wind mass loss rate
and enhanced atmospheric N and He abundances and reduced C and O
abundances. This indicates that it is already in an advanced stage of
core-hydrogen burning and has lost quite a lot of mass by its strong
wind and eruptions. This also points to a high initial mass, possibly
well in excess of 200\,\MSun\, \citep{2012ASSL..384....1H}.

We propose here the possibility that a third star in a relatively wide
orbit sparked the event, and discuss the physics of the ejection
process in this model.  Also if our model would finally not be the one
that explains all the chanacteristics of Eta Car, still and evolution
as discribed and modelled here is expected to happen not rarely in
nature.

Hierarchical triples are common among massive stars
\citep{2012arXiv1209.4638S}, and the orbital parameters required for
the proposed scenario to work appear to be rather common.  A
  triple origin of the great eruption has been suggested before by
  \cite{1998MNRAS.295L..59L}, in which an exchange in a dynamically
  unstable triple would have initiated the event, and by
  \cite{2003ChJAS...3..349K} who argued that an accretion disk around
  a tertiary neutron star would have sparked the great eruption.

According to our scenario, Eta Carinae was born a hierarchical
  triple, in which a tight inner binary star is orbited by a tertiary
  star in a wide orbit.  The inner binary system could have been
composed of a $M \apgt 90$\,\MSun\, primary star with an $m \simeq
10$---40\,\MSun\, companion in a rather tight $a_{\rm in} \simeq
1$\,AU orbit. By the time of the merger, the primary has lost some
10--20\,\MSun\, in a dense stellar wind.  A third companion in a
relatively wide orbit drives the inner binary, via the Lidov-Kozai
mechanism \citep{1962PSS..9..719L,1962AJ.....67..591K} to a state of
tidal evolution followed by coalescence.

A strong interaction between two stars has been demonstrated to
  be able to effectively drive a massive outflow similar to that
  observed in Eta Carinae
  \citep{2009NewA...14..539H,2011MNRAS.415.2020S}.  A similar model in
  which two main-sequence stars coalesce, was presented by
  \cite{2003AA...399..695K}.  Kato based his idea on the model by
  \citep{2003ApJ...582L.105S} for V838 Monocerotis \citep[see
    also][]{2006A&A...451..223T}.  Other binary merger models are
  those for V1309 Scorpius
  \citep{2011A&A...528A.114T,2014ApJ...786...39N} and V4332 Sagittarii
  \citep{2014arXiv1412.7822T}.  Due to the large distance, imaging of
  the ejecta in these cases is not feasable, except for V838 Mon which
  appears to be rather assymetric \citep{2003ApJ...582L.105S}.  The
  lightcurves of some of these potential merger events show
  characteristics very similar to that of the great eruption in Eta
  Carniae \citep{2010ApJ...709L..11K}.

The great eruptions in Eta Carina (but also of V838 Mon) appears to
have occurred near the pericenter of the current binary system
\citep{1996ApJ...460L..49D,2010ApJ...723..602K,2011MNRAS.415.2009S}.
As we will discuss in \S\,\ref{Sect:Discussion} this is consistent
with the model we discuss here.  In our case the pericenter passage
of the tertiary star initiates the merger of the inner binary system,
leading to the first great eruption of 1838.  The mass loss and
associated velocity assymmetry induced upon the binary merger causes
the very wide and eccentric outer orbit to shrink to its present size.
At the time of the second periastron passage in 1843 we expect the
merger product still to have been a quite extended star, such that
the new secondary collided with its outer layers, causing the large
1843 eruption.  The periastron passage of the outer companion at the
time of the merger, and therefore at the time of the great 1843
eruption, is thus a natural consequence of our simulations,
independently of the observations. 

If certain conditions are met, the resulting binary system will
resemble Eta Carinae. In this paper we discuss the conditions required
for the scenario to work, and argue that this model may provide a
plausible explanation for a number of curious observables for the Eta
Carinae system.

We will discuss the evolution of the triple system in three phases A,
B and C, which we relate to phases in the observed evolution of Eta
Car. The parameters describing the system are indicated with the
appropriate superscripts for the conditions upon birth (A), the
situation just before Roche-lobe overflow in 1838 (B), and after the
merger is completed (C).

\section{Eta Carinae before the great eruption (phase A)}\label{Sect:EtaCar1838_1843}

%(M=110\,\Rsun\, is +2\,\RSun compared to a 90MSun star)

At the age of 0.1---1 Myr, the $\sim 90$\,\MSun\, and 30\,\MSun\,
stars have radii as large as 15 --- 20\,\RSun\, and 4 --- 9\,\RSun\,
respectively. With an orbit separation of 70---400\,\RSun, these stars
will experience strong tidal interactions when the eccentricity grows
to $e \apgt 0.6$. As a result of such interaction the binary
circularizes, the rotation of the two stars synchronizes.  The tidal
energy that is dissipated in the stellar envelope gives rise to a
dramatic stellar wind from the primary star, until the two stars
eventaly engage in a common envelope
\citep{1984ApJ...277..355W,2006epbm.book.....E}.

A high eccentricity can be induced by a third body via the Lidov-Kozai
\citep{1962AJ.....67..591K} effect.  The rate at which this drives the
growth of the eccentricity of the inner orbit can be calculated from
secular perturbation theory \citep{2000ApJ...535..385F}. The time
scale of the secular growth can be comparable to the orbital period of
the inner binary. The growth of the inner eccentricity is particularly
rapid when the triple is barely dynamically stable
\citep{2008LNP...760...59M,2015arXiv150602039H}. It is in this range
of parameter space that the strongest interactions are expected, and
collisions may occur.

It is important to notice, as we will show in the next section, that
the tidal interaction may, already for many decades prior to the
merger, induce a much enhanced stellar wind from the primary star,
leading to a loss of more than 10\,\MSun\, prior to the merger (see
fig.\,\ref{Fig:dLdt}). We now will quantitatively study the evolution
of the triple-star model.

\subsection{Numerical integration of the triple before the merger 
  leading to a great eruption}\label{Sect:tripledynamics}

To quantify the orbital evolution of Eta Carina before the great
eruption in terms of the triple star model we follow the dynamical,
stellar and tidal evolution of the triple system.  We adopted the
Astronomical Multipurpose Software Environment \citep[AMUSE for
  short][]{2013CoPhC.183..456P}\footnote{The AMUSE source code is
  public and can be downloaded via \url{http:amusecode.org}. The
  scripts used to perform the simulations described here are also
  available via the same website.}) for this integration, by coupling
the various components and integrating the coupled system.

Gravity was solved using the {\tt Huayno} direct $N$-body integrator
\citep{2012NewA...17..711P}.  Energy in the dynamical simulations is
preserved to better than $1/10^{12}$ with respect to the initial
energy of the system, which is sufficient for a reliable integration
\citep{2041-8205-785-1-L3}. Tidal evolution was incorporated for the
two inner stars, but not for the tertiary, using the tidal dissipation
routine in {\tt AMUSE}, which is based on \citet[][see also
  \cite{2010ApJ...723..285H}]{1998ApJ...499..853E}.  For stellar
evolution we adopted the {\tt MESA} Henyey stellar evolution code
\citep{2011ApJS..192....3P}. All stars are assumed to be born with an
angular frequency of $\Omega = 2.6 \times 10^{-6} {\rm s}^{-1}$.  Each
time step we calculate the amount of energy dissipated in the tidal
interaction.

In the AMUSE script each of the three numerical solvers for stellar
evolution, tidal evolution and gravitational dynamics is run
subsequently, switching between them after every outer initial orbital
period and output was generated every 100 outer orbital periods. A
stopping condition was used to detect Roche-lobe overflow or a
dynamical instability \citep{2013A&A...557A..84P}.

In the event loop we first evolve all stars to the next time. In this
step we apply the mass loss calculated for the tidal interaction from
the previous step.  We then perform the tidal evolution and
recalculate the orbital elements and Cartesian coordinates of the
three stars. Here we assumed that the argument of periastron and the
mean anomaly are not affected by the tidal evolution.  We finally
integrate the gravitational dynamics of the triple system using the
direct $N$-body code.

In Table.\,\ref{Tab:DynamicalModels} we present an overview of initial
conditions adopted in this experiment.  We generally started with two
inner stars of $M_{\rm in}^{\rm A} = 110$\,\MSun\, and $m_{\rm
  in}^{\rm A} = 30$\,\MSun\, in a $a^{\rm A}_{\rm in}=1$\,AU orbit
with an eccentricity of $e^{\rm A}_{\rm in}=0.1$. The 30\,\MSun\,
outer star orbits the inner binary in an $a^{\rm A}_{\rm out} =
25$\,AU\, orbit with an eccentricity of $e^{\rm A}_{\rm out} = 0.2$ at
an inclination of $i^{\rm A}=90^\circ$. We also performed simulations
with a $M_{\rm in}^{\rm A} = 90$\,\MSun\, and a $M_{\rm in}^{\rm A} =
150$\,\MSun\, inner primary star or with an eccentricity of the outer
orbit of $e^{\rm A}_{\rm out} = 0.6$ and $a^{\rm A}_{\rm out} =
24$\,AU, with qualitatively gave the same results, although the moment
of Roche-lobe overflow varies somewhat.

\begin{table*}
  \caption{Initial conditions and results of the dynamical simulations
    with stellar evolution and tidal interaction (phase A). The
    secondary star of the inner binary ($m_{\rm in}^{\rm A}$) as well
    as the mass of the tertiary star ($m_{\rm out}$) are fixed to
    30\,\MSun. The inclination between the inner and the outer orbit
    was $i^{\rm A} = 90^\circ$. The first 6 columns give the other
    initial conditions, primary mass ($M_{\rm in}^{\rm A}$), fraction
    of tidal energy that drives the mass loss ($f$ between 0 and 1),
    orbital parameters (semi-major axis $a_{\rm in}^{\rm A}$ and
    eccentricity $e_{\rm in}^{\rm A}$) of the inner binary, followed
    by the parameters of the outer binary ($a_{\rm out}$ and $e_{\rm
      out}$).  The subsequent columns give the results of the
    simulations, starting with the moment of highest eccentricity
    ($t_{\rm max, e}$), the associated eccentricity reached ($e_{\rm
      in, max}$) and the maximal mass loss rate due to tidal
    interaction ($\dot{M}_{\rm max}$).  Subsequent columns give the
    time between reaching the maximal eccentricity and the moment of
    Roche lobe overflow ($dt_{\rm RLOF}$), the semi-major axis
    ($a_{\rm RLOF}$) and eccentricity ($e_{\rm RLOF}$) of the inner
    binary at RLOF, and the mass of the donor at this moment ($M_{\rm
      RLOF}$). The outer orbit, not listed in the table, is only
    slightly affected by the mass loss of the primary star
    \citep[see][for a qualitative discussion on
      this]{2013MNRAS.429L..45P}. }
  \label{Tab:DynamicalModels}
  \begin{center}
  \begin{tabular}{rlrr|llllllllrllllllllllll}
  \hline
 $M^{\rm A}_{\rm in}$ & $f_{\rm tidal}$ & $a^{\rm A}_{\rm in}$& $e^{\rm A}_{\rm in}$ & $a_{\rm out}$& $e_{\rm out}$ |
 &$t_{\rm max, e}$ & $e_{\rm in, max}$ & $\dot{M}_{\rm max}$& $dt_{\rm RLOF}$ & $a_{\rm RLOF}$& $e_{\rm RLOF}$&$M_{\rm RLOF}$ & Figure \\
\MSun& & AU & & AU & & Myr & & \Msun/yr & yr & \RSun & & \MSun \\
%[\MSun]& & [AU] & & [AU] & & [Myr] & & [yr] & [AU] & & [\MSun] \\

  \hline
 90 & 1 & 1.0 & 0.1 & 25.0 & 0.6 & 0.242 & 0.82 &0.591& 291 & 0.35 & 0.67 & 27.7 \\
110 & 1 & 1.0 & 0.1 & 25.0 & 0.2 & 0.255 & 0.87 &1.182& 127 & 0.64 & 0.64 & 70.4 \\
110 & 0.5 & 1.0 & 0.1 & 25.0 & 0.2 & 0.255 & 0.85 & 0.192 & 237 & 0.56 & 0.68 & 98.8 & see Figs.\,\ref{Fig:triple_sma_evolution}, \ref{Fig:dLdt} and \ref{Fig:CommonEnvelope}\\
110 &0.250& 1.0 & 0.1 & 25.0 & 0.2 & 0.256 & 0.86 &0.063& 177 & 0.57 &0.66 & 87.8 \\
110 &0.125& 1.0 & 0.1 & 25.0 & 0.2 & 0.256 & 0.85 &0.025& 225 & 0.59 &0.69 &105.0\\
110 & 0.0 & 1.0 & 0.1 & 25.0 & 0.2 & 0.255 & 0.84 &0.000& 118 & 0.78 & 0.72 & 110.0\\
110 & 0.5 & 1.0 & 0.1 & 25.0 & 0.6 & 0.256 & 0.86 &1.218& 177 & 0.57 & 0.66 & 87.8 \\
110 & 0.5 & 1.0 & 0.1 & 25.0 & 0.6 & 0.255 & 0.86 & 0.192 & 189 & 0.56 & 0.65 & 87.4 \\
110 & 0.5 & 1.0 & 0.1 & 24.0 & 0.6 & 0.006 & 0.79 &0.149& 297& 0.50 & 0.61 & 89.1 \\
110 & 0.5 & 1.0 & 0.1 & 20.0 & 0.2 & 0.086 & 0.64 &0.108& 683& 0.57 & 0.62 & 98.1 \\
110 & 0.5 & 1.0 & 0.1 & 22.0 & 0.2 & 0.111 & 0.47 &0.210& 588& 0.42 & 0.46 & 98.5 \\
%110 & 0.5 & 0.6 & 0.1 & 24.0 & 0.2 & 0.093& 0.58 &58876& 0.28 & 0.48 & 53.3 \\
%110 & 0.5 & 0.6 & 0.1 & 24.0 & 0.2 & 0.148& 0.68 & 883 & 0.30 & 0.67 & 26.8 \\
110 & 0.5 & 1.0 & 0.1 & 24.0 & 0.2 & 0.224& 0.86 &0.127& 228 & 0.61 & 0.68 & 89.9 \\
110 & 0.5 & 2.0 & 0.1 & 25.0 & 0.2 & 0.011& 0.90 &0.252& 213 & 0.53 & 0.65 & 85.0 \\
110 & 0.5 & 2.0 & 0.1 & 25.0 & 0.6 & 0.004& 0.90 &1.218& 103 & 0.44 & 0.59 & 77.4 \\
150 & 0.5 & 1.0 & 0.1 & 25.0 & 0.6 & 0.352& 0.81 &0.355& 125 & 0.62 & 0.61 & 123.8\\
  \hline
  \end{tabular}
  \end{center}
\end{table*}

In Tab.\,\ref{Tab:DynamicalModels} we notice that the time for
reaching Roche-lobe overflow for most of the calculated triples is
considerably longer than the classical Lidov-Kozai period, which is
caused in part by the coupled tidal and dynamical evolution, and in
part by the requirement that Roche-lobe overflow should ensue.
There may, however, still be a numerical effect, in particular in our
assumption that tidal evolution is instanteneous at regular time
intervals (of the outer orbital period) after which we progress this
information to the rest of the system under the assumption that the
argument of periastron and the orbital phase are not affected.  If the
outer orbit has a rather high eccentricity the collision tends to be
established much earlier than when the outer orbit has a relatively
small eccentricity. The choice of a high ($i^{\rm A} \sim 90^{\circ}$)
inclination is motivated by our follow-up study on the probable
conditions before the common envelope (see \S\,\ref{Sect:MCMC} in
which we discuss Markov-chain Monte-Carlo simulations to reconstruct
the parameters of the outer orbit prior to the common envelope).

\begin{figure*}
\begin{center}
\includegraphics[width=1.0\textwidth]{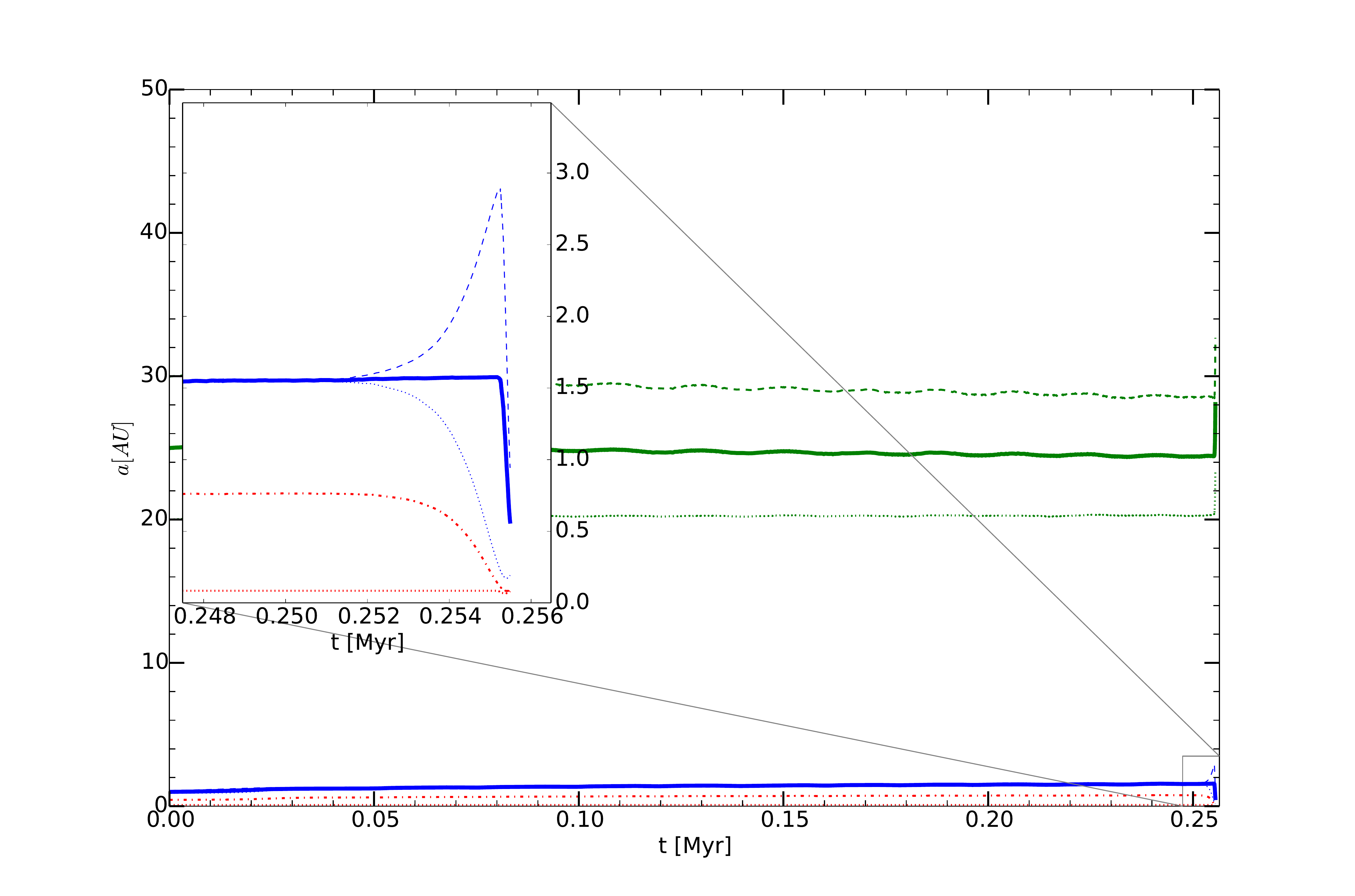}
\end{center}
\caption{Evolution of the orbital separation of an example triple
  system, described in the text.  The inner binary initially had a
  semi-major axis of $a^{\rm A}_{\rm in} = 1$\,AU and eccentricity
  $e^{\rm A}_{\rm in}=0.1$ with primary and secondary masses of
  $M^{\rm A}_{\rm in} = 110$\,\MSun\, and $m^{\rm A}_{\rm in} =
  30$\,\MSun. The outer orbit is inclined by $i^{\rm A}=90^\circ$ and
  has $a_{\rm out} = 25$\,AU and $e_{\rm out}=0.2$ with a tertiary
  mass of $m_{\rm out} = 30$\,\MSun. The top three curves (green) give
  the evolution of the outer orbital apocenter, semi-major axis and
  pericenter (from top to bottom). The bottom three blue curves (better
  visible in the inset to the top left) give the same parameters for
  the inner orbit. The dashed-dotted curve (red, second from bottom)
  gives an estimate of the instantaneous radius of the Roche lobe of
  the inner primary star \citep{2007ApJ...667.1170S}, and the bottom
  curve (dotted red) gives the radius of this star. The primary
  overfills its Roche lobe after 255,490\,yr.
\label{Fig:triple_sma_evolution}}
\end{figure*}

In Fig.\,\ref{Fig:triple_sma_evolution} we present the orbital
evolution of one of the calculations until the moment that the inner
primary fills its Roche-lobe.  
In this simulation the merger was initiated by a Lidov-Kozai cycle
that started about 0.255\,Myr after the birth the system and brought
the eccentricity to a maximum of 0.863. The subsequent tidal
interaction lasted for about 180\,yrs during which the semi-major axis
was reduced to $a_{\rm in} \simeq 0.563$\,AU with an eccentricity
$e_{\rm in} \simeq 0.66$. 

\subsection{Tidal energy dissipation and the possible origin of the Homonculus}\label{Sect:TidalLuminosity}

\begin{figure*}
\begin{center}
\includegraphics[width=1.0\textwidth,angle=-0.0]{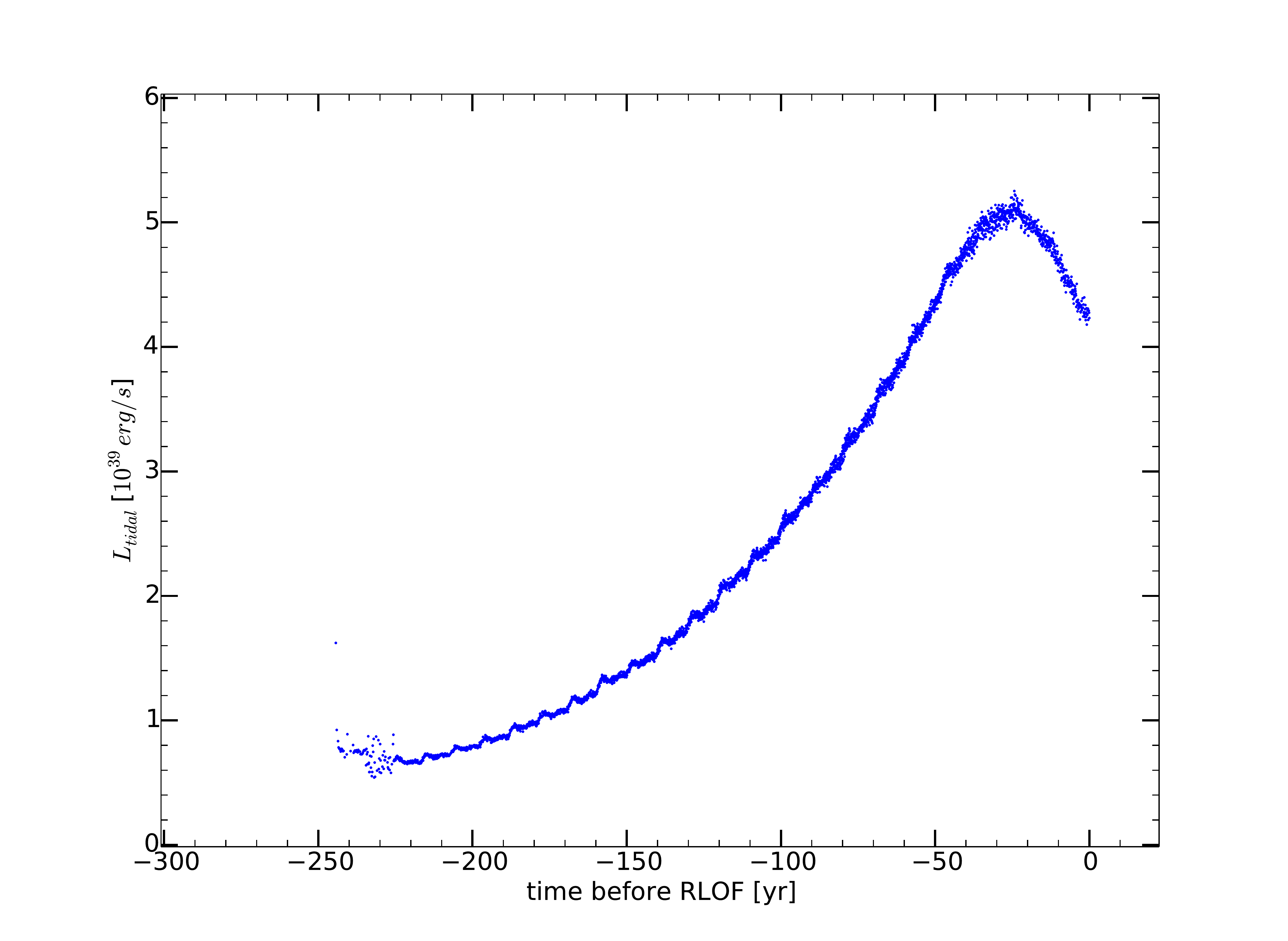}
\end{center}
\caption{Tidal energy dissipated in the donor star (of the inner
  binary system) prior to Roche-lobe overflow. Here Roche-lobe
  overflow occurs at $t=0$\,yr (which corresponds to 255,490\,yr after
  the birth of the system at the zero-age main-sequence).  The results
  are from the same simulation as presented for the orbital evolution
  in Fig.\,\ref{Fig:triple_sma_evolution}.  As explained in
  \S\,\ref{Sect:TidalLuminosity} we expect that due to the large tidal
  energy input the luminosity of the star to exceed the Eddington
  luminosity, leading to a huge continuum-drive wind mass loss of
  order 20\,\MSun\, in a few decades prior to the merger, which we
  suggest to have produced the Homonculus nebula.
\label{Fig:dLdt}}
\end{figure*}

During the $\sim 180$ year epoch before the common envelope much
energy is generated in the primary star by tidal
dissipation. Fig.\,\ref{Fig:dLdt} depicts the tidal energy generated
in this star as a function of time prior to the merger. The figure
shows that during the last 50 years prior to the merger the tidal
energy input into the star is of the order of $(4-6) \times
10^{39}$erg/s\,$ \simeq 1-1.5 \times 10^{6}$\,\LSun.

Already without this extra energy input the luminosity of this star is
close to the Eddington luminosity. \cite{2012ASSL..384..275O} have
argued that if due to an extra energy source temporarily the
luminosity exceeds the Eddington limit, very large
continuum-radiation-driven mass loss will set in, leading to an
eruptive phase, with an extremely high mass loss rate. This is because
a continuum-driven wind can in principle lead to mass-loss rates up to
the ``photon-tiring limit'', for which the entire luminosity is
expended in lifting the atmosphere. We therefore propose here that the
Homonculus, with a mass of about 20\,\MSun, was produced in an
eruptive phase of a few decades that resulted from the huge tidal
energy input in the star prior to the merger. In this phase the star
is very rapidly rotating and oblate. \cite{2012ASSL..384..275O} have
argued that in such a case the radiation-driven wind is much stronger
in the polar than in the equatorial regions of the star, leading to a
structure like the Homonculus. Therefore, the axis of the Homonculus is
expected to be along the rotation axis of the primary star, which is
perpendicular to the orbital plane of the inner binary.

\subsection{Evolution up to the merger}\label{Sect:evolutiontomerger}

The enhanced mass loss during the tidal event causes the primary star
to shrink from 18.4\,\RSun\, (before the strong tidal interaction),
down to 14.5\,\RSun\, with a rotational frequency of $\Omega \simeq
2.71 \times 10^{-5} {\rm s}^{-1}$ when the highest orbital
eccentricity is reached, which causes the star to swell to exceed
29.3\,\RSun\, at the equator, at which time the two stars engage in a
common envelope and coalesce.

The mass loss of the primary star prior to the merger, has caused the
outer orbit to expand slightly from $a_{\rm out} = 25$\,AU to $a_{\rm
  out} \simeq 28.1$\,AU but has a smaller effect on eccentricity of
the outer orbit $e_{\rm out} \simeq 0.17$ (starting from $e_{\rm out}
= 0.20$).

Each of our simulations of the evolution of the triple results in the
Roche-lobe overflow of the primary star. The inner binary is unstable
against mass transfer from the primary to the secondary star due the
small mass ratio (in most cases $q_{\rm in} \aplt 1/3$), which causes
rapid shrinking of the orbit. This, results in a common-envelope phase
leading to coalescence.  This coalescence between the inner two stars
is quite a violent event, the consequences are far reaching for the
entire triple. The energy released in such a merger process can
measure about 8--12$\times 10^{49}$\,erg \citep{2013ApJ...764L...6S},
which is sufficient to explain the great eruption
\citep{2012Natur.482..375R}.  In addition, the mass lost in the merger
process is likely to be asymmetric \citep{2006MNRAS.365....2M}, with
has quite specific consequences for the surviving binary system, which
after the merger is composed of the merger product and the tertiary
star.

To further quantify the merger event we perform smoothed particles
hydrodynamical simulations of the collision, which we describe in
\S\,\ref{Sect:SPH}.

\section{Eta Carinae during the great eruption (phase B)}\label{Sect:SPH}

We quantify the consequences of the coalescence by means of smoothed
particles hydrodynamics simulations of the merger process, again using
AMUSE. We are in particular interested in the amount of mass that is
ejected during the merger process, the direction in which this mass is
ejected, the resulting magnitude and the direction of the velocity
kick that is induced upon the merger product, the morphology of the
ejecta and the consequences for the orbital parameters of the outer
star.

The hydrodynamical simulations are performed using a binary system in
which the more massive primary star (with mass $M_{\rm in}$) and the
less massive secondary star (with mass $m_{\rm in}$) orbit each other
with semi-major axis $a_{\rm in}$ and eccentricity $e_{\rm in}$. For
now we ignore the tertiary star, because it hardly plays a role in the
merger process, and we will explore the effect of the merger on the
tertiary star further in \S\,\ref{Sect:MCMC}.

After the initial conditions are selected, the primary and secondary
stars are evolved using {\tt MESA}.  We first evolve the inner primary
star up to the moment that the dynamical evolution determines the
onset of Roche-lobe overflow (see \S\,\ref{Sect:tripledynamics}).  In
our calculations in which the primary was born as a 110\,\MSun\, star,
it is reduced through the tidal evolution mass-loss to about
87.6\,\MSun. By the time of Roche-lobe overflow at $t\simeq 255$\,kyr
it's radius measures 29.3\,\RSun, which is considerably larger than
its equilibrium radius of 18.4\,\RSun.  The tidal interaction has
further increased the star's rotational frequency to $\Omega \simeq
2.71 \times 10^{-5} {\rm s}^{-1}$ (see \S\,\ref{Sect:tripledynamics} for
details).  The companion star is evolved to the same age and we
adopted a rotational frequency of $\omega = 2.6 \times 10^{-6}
{\rm s}^{-1}$.

After the dynamical evolution of the triple (see
\S\,\ref{Sect:tripledynamics}) and upon Roche-lobe overflow both stars
in the inner binary are converted to a hydrodynamic particle
representation in order to be able to continue the evolution with an
smoothed particles hydrodynamics (SPH) code.  The conversion from a
1-dimensional stellar density and composition profile in the Henyey
code to the 3~dimensional particle representation as is used in an SPH
code was done with the method available in AMUSE, which is described
in \cite{2014MNRAS.438.1909D}.  All SPH particles have the same
mass. In this process the SPH particles are initially placed on a
grid, which is subsequently relaxed by running an SPH code for 10
dynamical time scales of the primary star.  We adopted the Gadget2 SPH
code \citep{2005MNRAS.364.1105S} for this relaxation process.

The two hydrodynamcal stars are subsequently placed in the
pre-determined binary orbit (see \S\,\ref{Sect:tripledynamics}
\$\,\ref{Sect:evolutiontomerger} and Tab.\,\ref{Tab:HydroModels}) and
the evolution of the system is continued.  For the latter we adopted
the SPH code {\tt Fi} \citep{2004A&A...422...55P}.  Changing codes in
AMUSE is self-consistent and easy: it requires changing a single line
in the Python-AMUSE script. Running the relaxation of the stars with
one SPH code and the subsequent gravitational/hydrodynamical evolution
of the system with another is therefore motivated by performance and
applicability of the particular code for that specific task. The {\tt
  Fi} SPH code is, in this case, better suited for simulating the
merger process, whereas {\tt GadGet2} was already adopted and
extensively tested in \cite{2014MNRAS.438.1909D} for converting Henyey
stellar evolution models to SPH realizations.

\begin{figure}
\begin{center}
\includegraphics[width=0.5\textwidth,angle=-0.0]{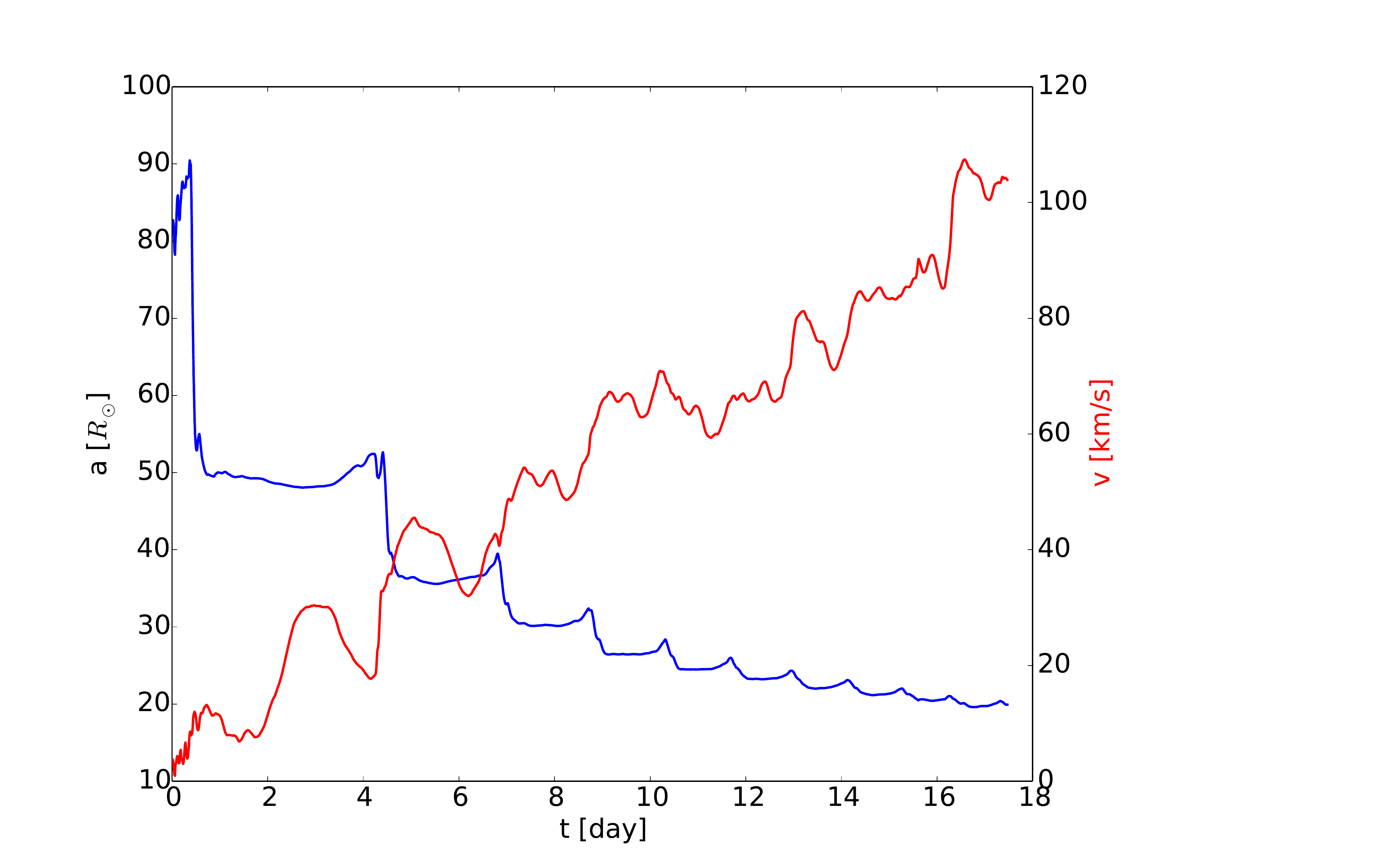}
\end{center}
\caption{Orbital evolution of the inner binary until it merges (blue,
  axis to the left), and the velocity of the bound mass (red, axis
  to the right). These curves are the result of an SPH simulation
  using two stars of $M_{\rm in}^{\rm B} = 70$\,\MSun, and $m_{\rm
    in}^{\rm B}=30$\,\MSun\, at an age of 1.0\,Myr in a $a_{\rm
    in}^{\rm B}=77$\,\RSun\, and with eccentricity $e_{\rm in}^{\rm
    B}=0.87$.  The $m_{\rm out} = 30$\,\MSun\, tertiary star has
  highly eccentric $e_{\rm out}=0.9$ orbit of $a_{\rm out} =
  100$\,AU\, around the inner binary. The relative inclination and
  argument of pericenter are both $i^{\rm B} = 90^\circ$.  The
  Lidov-Kozai cycles induced by this outer star cause the eccentricity
  of the inner binary to grow to $e_{\rm in}^{\rm B}=0.87$ (see phase
  A), upon which the two inner stars coalesce.  The merger induces a
  kick of about $v_{\rm kick} = 81.5$\,km/s onto the merger product,
  which resulted from the asymmetric ejection of the common-envelope
  of the inner binary. This kick was directed towards pericenter of
  the original inner binary.  (Note, that this system is different
  than used in Fig.\,\ref{Fig:dLdt}).
\label{Fig:CommonEnvelope}}
\end{figure}

We adopted the parameters for the inner orbit just before the merger
from the dynamical simulations described in
\S\,\ref{Sect:tripledynamics} (and Tab.\,\ref{Tab:HydroModels}).  From
the moment of first contact at pericenter, the two stars orbit each
other up to a dozen times before the merger is complete. In
fig.\,\ref{Fig:CommonEnvelope} we present the evolution of the
semi-major axis of one of these models.

In some of our simulations the binary did not result in a merger
within 100 days, after which we stopped them.  The latter cases could
eventually result in a merger, but the parameters would be different
due to the continuing Kozai-Lidow cycles and stellar evolution, which
we ignored in the SPH simulations, because we omitted the tertiary
star in the hydrodynamical simulations.

The results of the SPH simulations are analyzed using HOP
\citep{1998ApJ...498..137E} with the saddle-point density cut-off set
to the mean density and the outer density threshold to 5\% of the mean
density.  We continue running the SPH simulation until only one single
blob of particles remains.  After the first detection of a single
blob, we continue to run the hydrodynamics for an orbital period of
the outer star to assure that the single object really remains
integer.  This single blob is considered to be the merger product, and
the residual material is identified as the ejecta in the merger
process.

\begin{table*}
  \caption{Initial conditions and results for the smoothed particles
    hydrodynamics simulations.  The first column gives the number of
    SPH particles for the primary star.  The SPH particles in the
    simulations for the primary and secondary star have the same mass.
    and the total number of SPH particles is The following columns
    give the age of the system at the moment of Roche-lobe overflow,
    and the masses of the primary ($M^{\rm B}_{\rm in}$) and secondary
    ($m^{\rm B}_{\rm in}$) stars at that moment.  In the fifth and
    sixth columns we give the initial orbital separation $a^{\rm
      B}_{\rm in}$ and eccentricity $e^{\rm B}_{\rm in}$ of the binary
    at the moment of Roche-lobe overflow.  The last three columns give
    the duration of the merger process, the final total mass in the
    merger product ($M^{\rm C}$) and its velocity $v_{\rm kick}$ as a
    result of the asymmetric mass ejection during the merger process.
    In the SPH simulations we ignore the effect of the tertiary star,
    because the effect of Lidov-Kozai cycles on the inner orbit is
    negligible on the short time scale of the merger.  }
  \label{Tab:HydroModels}
  \begin{center}
  \begin{tabular}{rlllllllllllllllllllllllll}
  \hline
$N_{\rm SPH}$&$t_{\rm RLOF}$&$M^{\rm B}_{\rm in}$&$m^{\rm B}_{\rm in}$& $a^{\rm B}_{\rm in}$& $e^{\rm B}_{\rm in}$ &$t_{\rm mrg}$&$M^{\rm C}$&$v_{\rm kick}$ & Figure\\
$\times 1000$ &Myr&\multicolumn{2}{c}{\MSun}&\RSun& &day& \MSun & km/s\\
  \hline
 10 & 0.96& 70 & 30 &  77& 0.87 & 54.3 & 77.3 & 193 \\ 
 50 & 0.96& 70 & 30 &  77& 0.87 & 17.3 & 68.8 & 81.5  & see Fig.\,\ref{Fig:CommonEnvelope}\\
100 & 0.49& 80 & 20 &  77& 0.87 & 26.4 & 83.1 & 70.6\\
 10 & 0.49& 80 & 20 &  77& 0.87 & 21.2 & 83.8 & 58.6\\
 50 & 0.49& 80 & 20 &  77& 0.87 & 15.5 & 83.2 & 61.7 \\
 50 & 0.13& 90 & 10 &  77& 0.87 & 27.4 & 90.1 & 211  \\
  2 & 0.49& 80 & 20 &  77& 0.87 &  7.5 & 86.3 & 34.8 \\
  4 & 0.49& 80 & 20 &  77& 0.87 & 11.1 & 84.7 & 12.6 \\
  8 & 0.49& 80 & 20 &  77& 0.87 & 13.0 & 83.7 & 30.3 \\
 16 & 0.49& 80 & 20 &  77& 0.87 & 22.3 & 83.6 & 52.7  \\
 20 & 0.49& 80 & 20 &  77& 0.87 &  7.1 & 78.0 & 56.5  \\
 32 & 0.49& 80 & 20 &  77& 0.87 & 7.9 & 83.6 & 56.9 \\
 32 & 0.49& 80 & 20 &  77& 0.87 & 7.9 & 83.4 & 56.9 \\
 64 & 0.49& 80 & 20 &  77& 0.87 & 8.5 & 77.5 & 47.7 \\
128 & 0.49& 80 & 20 &  77& 0.87 & 8.9 & 77.8 & 54.5 & see Figs\,\ref{Fig:SPH_xy_post_merger} \& \ref{Fig:3DSPH_xy_post_merger} \\
256 & 0.49& 80 & 20 &  77& 0.87 & 5.8 & 78.1 & 60.9 \\
 16 & 0.47&124 & 30 &  77& 0.87 & 2.8 & 138 & 25.1 \\
 32 & 0.47&124 & 30 &  77& 0.87 & 3.2 & 137 & 18.6 \\
 16 & 1.82& 88 & 30 & 121& 0.92 & 9.5 & 100 & 12.3 \\
 32 & 1.82& 88 & 30 & 121& 0.92 & 7.8 & 101 & 14.5 \\
%??  8000 & 0.26& 90 & 30 & 121& 0.88 &14.00 & 105.9 & 24.4 \\
  \hline
  \end{tabular}
  \end{center}
\end{table*}

\begin{figure}
\begin{center}
\includegraphics[width=0.5\textwidth,angle=-0.0]{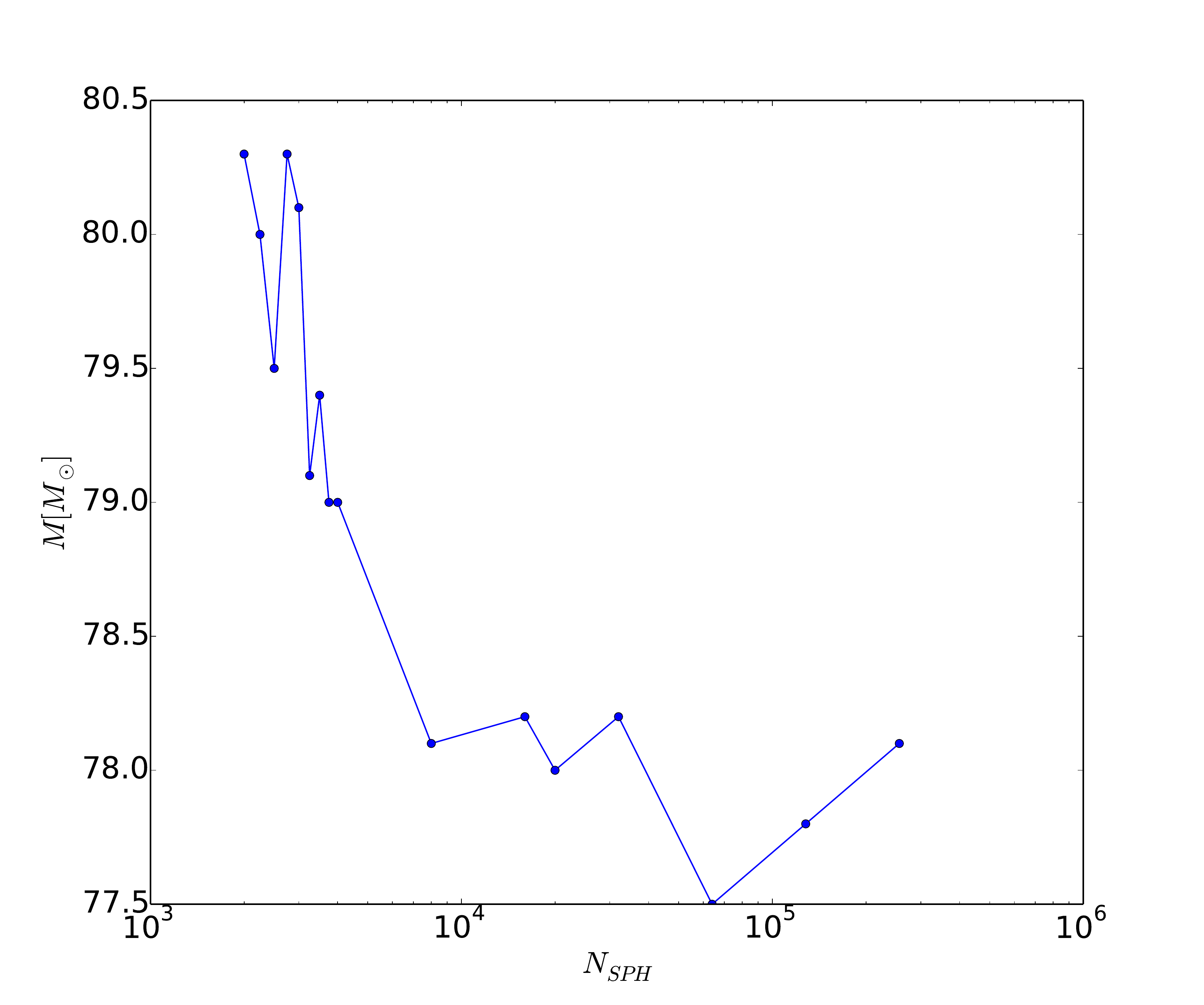}
\end{center}
\caption{Mass of the merger product as a function of the number of SPH
  particles for the primary star. Each bullet point is one simulation
  using the same initial conditions but with varying number of SPH
  particles. The initial conditions at the moment of RLOF were $M^{\rm
    B}_{\rm in} = 90$\,\MSun, $m^{\rm B}_{\rm in} = 30$\,\MSun,
  $R_{\rm in} = 15$\,\RSun, $a^{\rm B}_{\rm in} = 55$\,\RSun\, and
  $e^{\rm B}_{\rm in}= 0.845$.  The mass of the merger product
  converges to about $M^{\rm C} = 78$\,\MSun\, which indicates at
  about 12\,\MSun\, was ejected in the merger process.
\label{Fig:SPHconvergence}}
\end{figure}

We performed several convergence tests, one is presented in
fig.\,\ref{Fig:SPHconvergence}. Here we show the mass of the merger
product as a function of the number of SPH particles.  The mass of the
merger product seems to converge at $\apgt 10^4$ SPH particles for the
primary star; lower resolution simulations systematically
underestimate the amount of mass ejected in the merger process. We
performed similar analyses for the time or the merger, the velocity of
the ejecta and the merger product, and for the direction and
morphology of the ejected material. Each study shows convergence for
$\apgt 10^4$ SPH particles.

\subsection{The stellar collision merger product}

We realize that the number of simulations we performed is rather
limited, and can hardly be considered a complete coverage of parameter
space, but we think that our simulations are representative for the
physical process we try to describe.

\begin{figure}
\begin{center}
\includegraphics[width=0.5\textwidth,angle=-0.0]{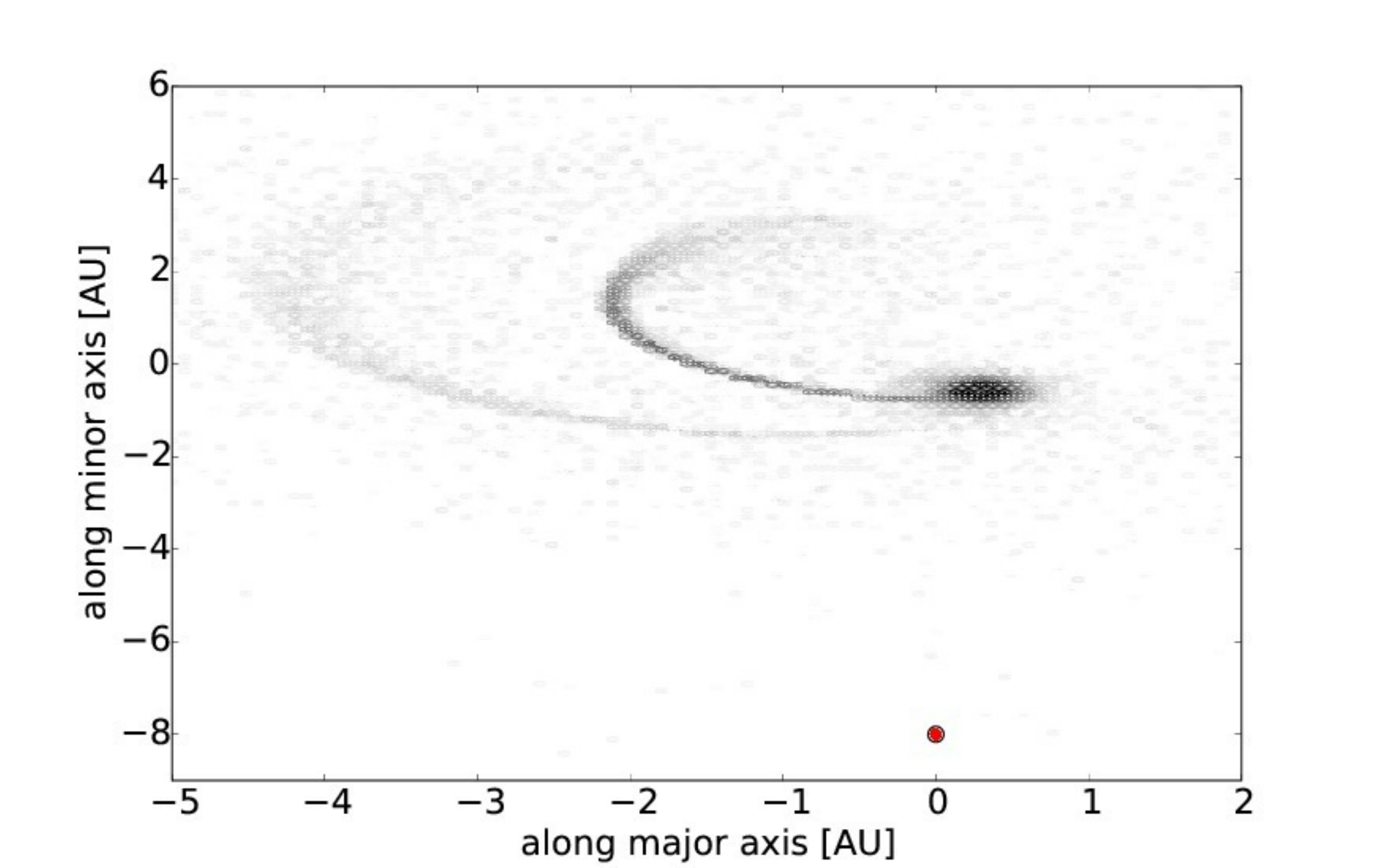}
\end{center}
\caption{Projected density distribution (in gray) of the gas ejected
  in the merger process at the moment that the two stars in the inner
  binary have merged to one. The origin is in the original center of
  mass of the inner binary system. The two massive loops of ejected
  gas (to the left of the origin) are quite characteristic for the
  merger process. In Fig.\,\ref{Fig:3DSPH_xy_post_merger} we present a
  more appealing picture of the same merger.  At this moment the
  tertiary star will be near pericenter at the indicated position (red
  bullet) moving away from the reader.  The hydrodynamical simulations
  are taken 8.8 days after the merger between a $M^{\rm B}_{\rm in} =
  80$\,\MSun\, star and a $m^{\rm B}_{\rm in} = 20$\,\MSun\, star in
  an $a^{\rm B}_{\rm in} = 77$\,\RSun\, orbit with an eccentricity of
  $e^{\rm B}_{\rm in} = 0.87$ (see Tab.\,\ref{Tab:HydroModels}).
\label{Fig:SPH_xy_post_merger}}
\end{figure}

We realize that we should not interpret the result of the SPH
simulations statistically, but by a lack of understanding the initial
parameter space we discuss them as a canonical ensemble. On average
the merger takes $14 \pm 12$ days during which $18.3 \pm 4.3$\,\MSun\,
is lost from the two inner stars.  This mass is ejected by the violent
in-spiral of the original inner binary system in a $\sim 70^\circ$
shell centered around the direction of the apocenter of the inner
orbit.  In Fig.\,\ref{Fig:SPH_xy_post_merger} we present the spatial
morphology of the out flowing gas 8.8 days after the merger of our
hydrodynamical simulation in which an 80\,\MSun\, star merges with a
20\,\MSun\, star from an 77\,\RSun\, orbit with an eccentricity of
0.87.  The velocity of the ejected material is about 400---500km
\,s$^{-1}$.

Earlier calculations arrive at similar conclusions regarding the
velocity and direction of the ejecta
\citep{1961Obs....81...99T,1978ApJ...219..498W}.  Due to conservation
of momentum, the merger product receives a kick of $v_{\rm kick} =
76.4 \pm 53.9$\,km/s in the plane and roughly in the direction of
pericenter in the inner orbit.

Upon its subsequent pericenter passage, some 5.5 years later, the
outer star plunges through the bloated merger product, giving rise to
the 1843 outburst.

We have illustrated the chain of events that may have led to the great
eruption of Eta Carinae. We will now demonstrate how the model may
satisfactorily explain its observed orbital parameters by means of
Markov-chain Monte Carlo simulations of the evolution of the triple
through the merger process.

\section{Reconstructing the triple topology during the giant eruption}\label{Sect:MCMC}

The parameter space for triples that engage in an evolution similar to
that described in the previous section is rather large and simulating
the secular evolution of the triple system and the subsequent merger
are too expensive in terms of computer time to fully map parameter
space. Ideally, we would like to perform a exhaustive search of
parameters using our model in order to constrain the initial
conditions and the events that led to the currently observed system,
but that is not realistic. Instead, we construct a model of the
consequences of the merger by making a few simplifying assumptions,
and adopt a Markov-chain Monte Carlo approach to map parameter space
and constrain the pre-merger parameters of the triple system.

\begin{figure}
\begin{center}
\includegraphics[width=0.5\textwidth,angle=-0.0]{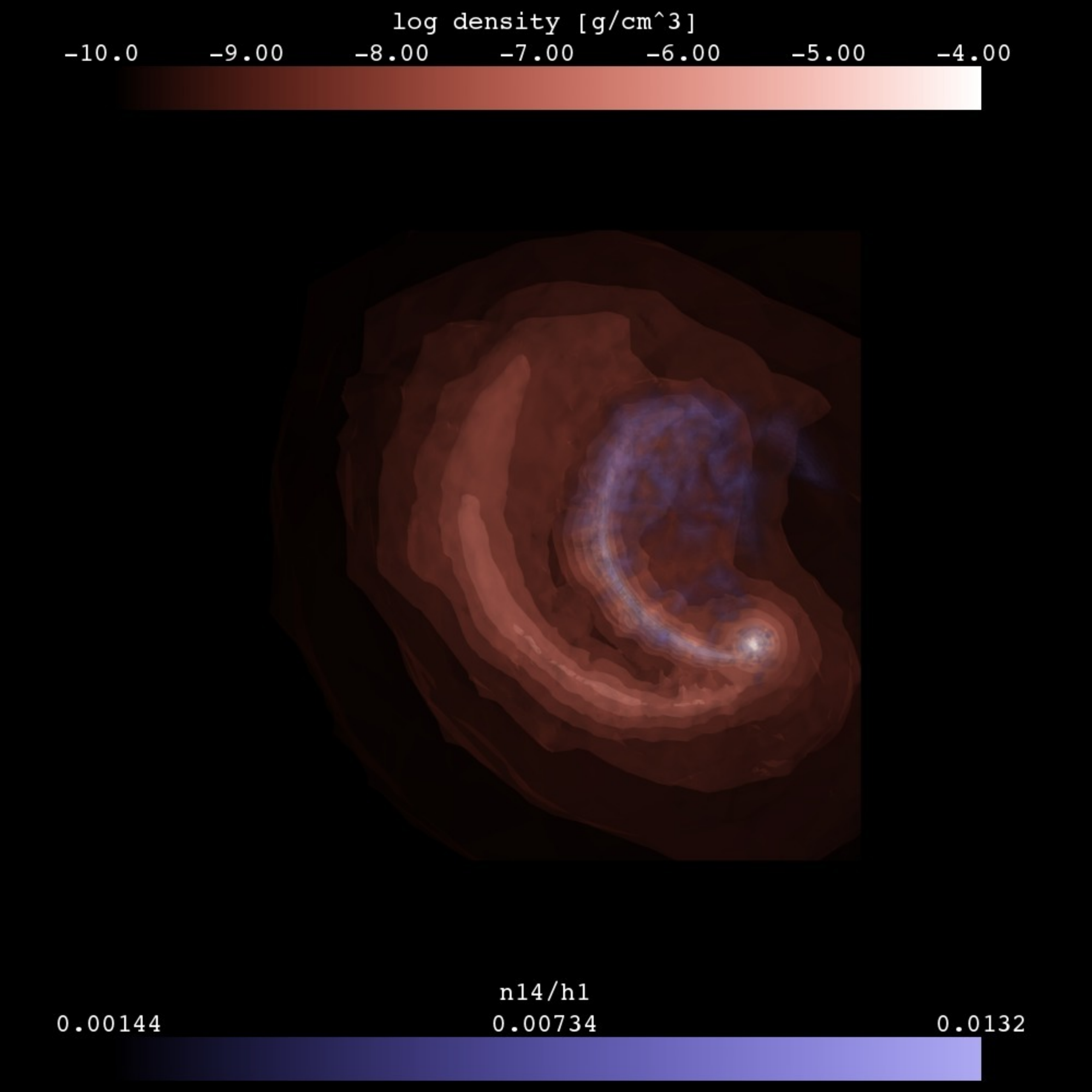}
\end{center}
\caption{Same as Fig.\,\ref{Fig:SPH_xy_post_merger} but then visually
  more attractive and in a rectangular frame of about 12\,AU at each
  side. The projected density is given in red, and the
  Nitrogen/hydrogen ratio is presented blue.
\label{Fig:3DSPH_xy_post_merger}}
\end{figure}

This approach is separate and independent of the earlier calculations
(see \S\,\ref{Sect:EtaCar1838_1843} and \S\,\ref{Sect:SPH}) and there
is no guarantee that the results of the Monte Carlo simulations
provide reasonable conditions prior to the merger. The fact that the
model provides reasonable conditions bolsters our confidence in the
proposed scenario for the giant eruption of Eta Carinae.

One of the assumptions we will make is that the time scale for the
merger is short compared to the orbital period of the outer star in
the triple system ($\sim 12$\,yr).  The SPH simulations of the merger
process could be viewed as a guideline for these time scales, which
turn out to be $\aplt 25$\,days (see \S\,\ref{Sect:SPH},
Fig.\,\ref{Fig:CommonEnvelope} and Tab.\,\ref{Tab:HydroModels}).  The
merger itself is therefore perceived by the outer star as an impulsive
event, in which the inner binary merger loses mass and receives a
velocity kick due to the asymmetry with which the mass is ejected.
For the duration of the merger we further assume that the inner object
(the merging binary star) is perceived by the outer star as a point
mass.

The kinematics of the gas that is ejected in the merger process is
less well constrained.  The initial outer orbital period exceeds the
initial inner orbital period by at least two orders of magnitude and
it takes some time before the mass shell has passed the outer orbit.
As we will discuss later, the merger event is likely to take place
when the outer star is near its pericenter, which ensures that the
time needed for the ejected mass shell to escape the (by then) binary
system is small (30---70\,days) compared to the orbital period of the
outer star ($\apgt 10$\,years).

The time scale for the outer star to re-approach pericenter is about
an orbital period of the post-merger binary system. This gives rise to
a second outburst, some time after the 1838 merger event, which lasted
several tens of days. This second outburst is then associated with the
companion star plunging through the bloated merger product.

\subsection{Reconstructing the 1983 event}

We solve Kepler's equations for a two body system while taking the
mass loss and the impulsive kick to the merger into account. Similar
calculations have been performed in relation to supernovae in binary
\citep{1983ApJ...267..322H,1998A&A...330.1047T} and triple systems
\citep{2012MNRAS.424.2914P}.
 
We adopt the Metropolis-Hastings algorithm \citep{HASTINGS01041970} to
reconstruct the initial pre-merger parameters of the triple system by
using the currently observed orbital parameters for Eta Carinae as
objective (see Tab.\,\ref{tab:EtaCar}). The fixed parameters are the
masses of the two stars in the binary after the merger ($M^{\rm
  B}_{\rm in} + m^{\rm B}_{\rm in} \equiv M^{\rm C} = 90\,\MSun$ for
the merger product and $m_{\rm out} = 30\,\MSun$ for its companion).

\begin{figure*}
\begin{center}
\includegraphics[width=1.0\textwidth]{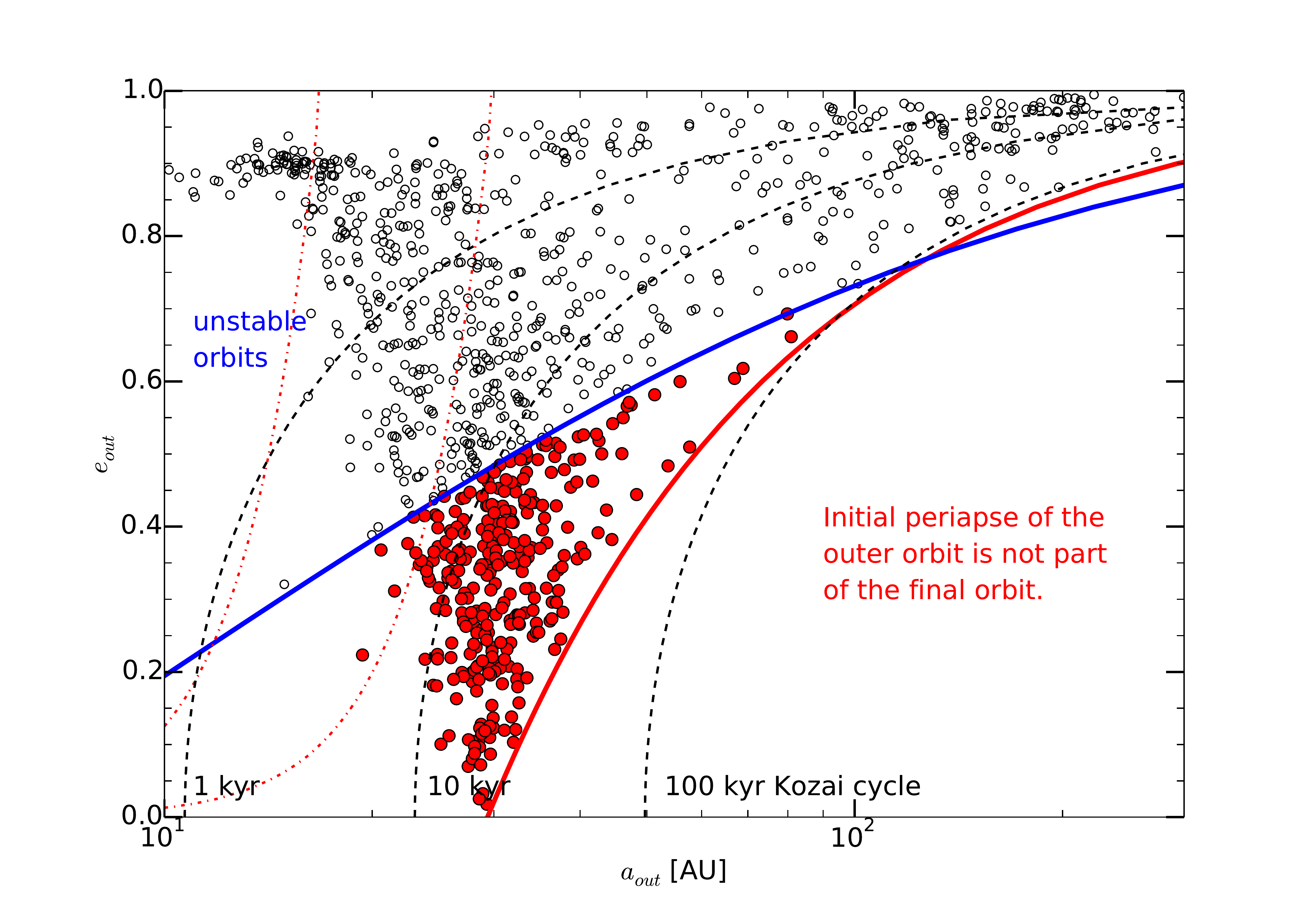}
\end{center}
\caption{Range of solutions for the progenitor outer binary that can
  result in consistent currently observed orbital parameters of Eta
  Carinae.  We adopted the inner stars of $M^{\rm B}_{\rm in} =
  90$\,\MSun\, and $m^{\rm B}_{\rm in} = 30$\,\MSun\, in a circular
  inner orbit of $a^{\rm B}_{\rm in} = 1$\,AU, the outer star was
  $m_{\rm out} = 30$\,\MSun.
%%%%%%%%%%%%%%%%%%%%%%%%%%%%%%%%%%%%%%%%%%%%%%%%%%%%%%%%%%%%%%%%%%%%%%%
  Parameter space for viable solutions to the Lidov-Kozai induced
  merger.  The dashed curves give an estimate of the time scale for
  Kozai cycles, from left to right 1kyr, 10kyr and 0.1Myr,
  respectively.  
%%%%%%%%%%%%%%%%%%%%%%%%%%%%%%%%%%%%%%%%%%%%%%%%%%%%%%%%%%%%%%%%%%%%%%%
  The red dash-dotted curves indicate parameters for which the
  pericenter distance was reduced due to the Lidov-Kozai effect by
  more than 0.1\% and 0.01\% (from left to right) of the radius of the
  secondary between two subsequent pericenter passages
  \citep{2000ApJ...535..385F}.
%%%%%%%%%%%%%%%%%%%%%%%%%%%%%%%%%%%%%%%%%%%%%%%%%%%%%%%%%%%%%%%%%%%%%%%
  The red solid curve gives the minimum distance in the orbit of the
  outer star such that it is part of the currently observed orbit of
  Eta Carinae: $p = R_{\rm sum}/a_{\rm EtaCar}(1+e_{\rm
    EtaCar})$. Here $R_{\rm sum}$ is the sum of the radii of the inner
  two stars. This gives a minimal orbital separation for which an
  impulsive event can lead to a viable observed orbit.  Initial outer
  orbits below this curve have no part of the orbit that is consistent
  with the currently observed orbit of Eta Carinae, and therefore do
  not lead to possible solutions.
%%%%%%%%%%%%%%%%%%%%%%%%%%%%%%%%%%%%%%%%%%%%%%%%%%%%%%%%%%%%%%%%%%%%%%%
  Systems above the blue solid curve are dynamically unstable at the
  moment of Roche-lobe overflow \citep{2008LNP...760...59M}, and
  therefore do not provide a possible solution.  Only those systems
  (red bullets) between the solid blue and the solid red curve can
  lead, through a merger event, to a system with parameters consistent
  with the currently observed orbit of Eta Carinae (see
  Tab.\,\ref{tab:EtaCar}).
%%%%%%%%%%%%%%%%%%%%%%%%%%%%%%%%%%%%%%%%%%%%%%%%%%%%%%%%%%%%%%%%%%%%%%%
  The circles give the result of the Monte-Carlo simulations aiming to
  reproduce the current orbital parameters ($a_{\rm EtaCar}$ and
  $e_{\rm EtaCar}$) for Eta Carinae.  The open symbols give
  dynamically unstable solutions, whereas the bullets give stable
  solutions.
\label{Fig:MonteCarlo}}
\end{figure*}

The free parameters in the Markov chain are the total mass of the
inner two stars before the merger, magnitude and direction of the
velocity imparted to the merger product and the orbital parameters of
the outer orbit ($a_{\rm out}$, $e_{\rm out}$, orbital phase and
relative inclination $i$) at the moment of merging.  We search
parameter space using the {\tt EMCEE} \citep{2013PASP..125..306F}
implementation for a Markov-chain Monte Carlo simulations.  The
solution is degenerate, and therefore we run the Monte Carlo code
several times in order to construct an ensemble of possible
solutions. Each Monte Carlo realization is performed with 10 workers,
a burn-in of 10 generations and with 200 subsequent iterations, which
is sufficient for the solution to converge to within 1\% of the
observed parameters for Eta Carinae. In Fig.\,\ref{Fig:MonteCarlo} we
present the result of the Monte Carlo simulations, in particular the
semi-major axis and eccentricity of the outer orbit prior to the
merger event.  This is supposedly the moment after the dynamical
simulations described in \S\,\ref{Sect:tripledynamics} (phase A) but
before the onset of the common-envelope evolution discussed in
\S\,\ref{Sect:SPH} (phase B).

The currently observed orbital parameters for the pre-great eruption
binary system are best reproduced when the inner binary was orbited by
another star in a relatively wide $a_{\rm out} = 20$---100\,AU orbit
with an eccentricity of $e_{\rm out} \aplt 0.7$.  A total of
12---23\,\MSun\, was lost, preferentially in the direction towards
apocenter along the semi-major axis of the inner binary, inducing a
velocity kick to the merger product of $v_{\rm kick} = 53 \pm 8$\,km/s
(with a minimum of 40km/s) in the orbital plane of the inner binary
and pointing to its pericenter ($\phi = 180^\circ \pm 9^\circ$) and
almost perpendicular to the plane of the outer binary ($\theta =
91^\circ \pm 21^\circ$).  This latter angle is interpreted as the
relative inclination between the inner and the outer orbit. (Note that
this finding motivated us in \S\,\ref{Sect:tripledynamics} to adopt
inclinations $i^{\rm A} = 90^\circ$.) The required mass loss and
velocity kick are consistent with our earlier results from the
hydrodynamical calculations presented in Table\,\ref{Tab:HydroModels}
($18.3 \pm 4.3$\,\MSun\, and $v_{\rm kick} = 61 \pm 51$\,km/s, see
\S\,\ref{Sect:SPH}).  The orbit of the outer star has, at the moment
of the merger, a mean anomaly of ${\cal M} = 0.061 \pm 0.035$, which is
just barely past pericenter.

Such a highly inclined outer orbit induces Lidov-Kozai
\citep{1962AJ.....67..591K} cycles onto the inner orbit with a period
of $\apgt 1$\,kyr, and drives the eccentricity of the inner orbit to
exceed $\sim 0.97$.  For tight inner orbits, $a^{\rm A}_{\rm in} =
0.30$---0.6\,AU, these cycles can last for as long as 1~Myr. Long
periods (and tight inner orbits) are consistent with the observed age
of Eta Carinae, but are possibly less suitable for inducing a
collision via Lidov-Kozai cycles, because the variation in the orbital
eccentricity (and therefore in the pericenter distance) per inner
revolution is relatively small.  However, as we discussed in
\S\,\ref{Sect:tripledynamics}, the response of the donor to the tidal
evolution may be quite important for the merger process.  In our
simulations in \S\,\ref{Sect:EtaCar1838_1843} we discussed the growth
of the tidally heated inner primary star, which is likely to initiate
the merger.

A collision is more probable when the inner orbit is larger 1---2\,AU,
because this leads to large changes in the pericenter distance within
one inner orbital revolution (see Fig.\,\ref{Fig:MonteCarlo}).
Considering the large uncertainty in the age determination of Eta
Carinae, we see sufficient leeway to argue that the system is
considerably younger ($\aplt 0.4$\,Myr) than 1\,Myr.  We therefore
have a slight preference for relatively wide inner orbits, which
motivated us to perform most of the simulations in
\S\,\ref{Sect:tripledynamics} with $a_{\rm in} \simeq 1$\,AU.

\section{Discussion}\label{Sect:Discussion}

We have explored the possibility that the great eruption of Eta
Carinae was initiated by a tidal interaction followed by the common
envelope merger of the two stars of the inner binary in a hierarchical triple
system.  We support theoretical arguments with numerical simulations
of various stages of the process. We find consistency with a number of
observables, including various phases of fast and slow mass loss, the
brightening of the object prior to the great eruption, and the events
are consistent with the current orbital parameters of the surviving
binary system. Encouraged by the successes, we will discuss some of
these advantages of our model qualitatively and also raise some
concerns.

\subsection{The brightening of Eta Carinae prior to the great eruption}\label{Sect:Brightening}

Our merger model can, at least qualitatively, account for the gradual
increase in brightness of Eta Carina in the decades preceding the
large outbursts of 1838-1843.  During the first half of the nineteenth
century (and possibly even in the two centuries before that) the
visual brightness of Eta Carina had increased to magnitudes 4 or
brighter, from a quiescent value of $\sim 8$\, magnitudes, which it
assumed again after year 1900 (although after 1940 it brightened again
to about visual magnitude 5, which has been ascribed to dust
destruction and expansion of a dust envelope
\cite{2012ASSL..384....1H}). This large visual luminosity increase
cannot be solely explained by a bolometric correction effect of a star
with an expanding photospheric radius at constant bolometric
luminosity: this may explain at most perhaps an increase of 2.5 to 3
visual magnitudes. Therefore it appears that the bolometric luminosity
of Eta Carina already increased considerably in the centuries
preceding the giant outburst, which means that somehow extra energy
must have been generated in its interior during this period. Hydrogen
fusion in stellar interiors tends to be highly stable and
self-limiting \cite[hydrogen-fusing stars have an in-built feedback
  ``safety-valve'': a slight overproduction of nuclear energy leads to
  increase in core temperature, leading to increased core gas
  pressure, which will make the core expand and adiabatically cool,
  quenching the increased rate of fusion][]{2006epbm.book.....E}. It
therefore seems highly unlikely that this increase would be due to
increased nuclear fusion in the interior as has, for example, been
suggested by \cite{2012ASSL..384..275O}. (It should be noticed,
however, that in stellar interiors where radiation pressure plays a
dominant role the feedback loop of fusion has not yet been well
studied).

We argue here that the increased bolometric luminosity preceding the
large outburst is due to tidal friction produced by the in-spiralling
companion star. In \S\,\ref{Sect:EtaCar1838_1843} we estimate the
energy that can be generated by this friction, which shows that it
may, in principle, lead to an energy release comparable to the
luminosity increase preceding the giant eruption.  We argued there
that since the unperturbed star is already very close to its Eddington
luminosity, a large increase of its luminosity due to tidal friction
is in itself sufficient to cause a phase of eruptive mass-loss that
could have produced the Homonculus.

\subsection{The surface abundances of Eta Carinae}\label{Sect:Discussion.sabundancy}

Eta Carinae has an enhanced atmospheric N and He abundances and
reduced C and O abundances \citep{2012ASSL..384....1H}.  The surface
abundance of a star can be calculated using our adopted Henyey stellar
evolution code (see \S\,\ref{Sect:tripledynamics}).  We perform this
calculation by measuring the abundances in the outer-most shell in the
Henyey code for the 110\,\MSun\, zero-age star in the triple
calculation of \S\,\ref{Sect:tripledynamics} at the moment of RLOF.
Upon RLOF the mass of the star was reduced due to the tidally induced
stellar wind to about 87.6\,\MSun.  We compare the surface abundance
with the resulting surface composition of the merger product of our
hydrodynamical simulations in \S\,\ref{Sect:SPH}.  For the latter we
measure the abundances of the outer most 1\% of the SPH particles in
the simulations with $N_{\rm SPH} = 128,000$. The former measurement
has no formal error, whereas the result of the SPH calculations has a
Poissonian error.

The merger process leads to a slight enhancement in the helium
abundance, from a relative fraction of 0.28 in the Henyey code to
$0.295 \pm 0.018$ after the merger simulation; a relative increase of
5.4\%. Nitrogen however, is enhanced through the merger process by a
factor 3.7 (from 0.00101 to $0.0047 \pm 0.0034$).  Our measured
nitrogen enhancement however, is still about a factor of 2 smaller
than the observed enhancement \citep{2005AAS...20711411M}.  Oxygen in
our simulations is depleted by 19\% (from 0.0094 to $0.0076 \pm
0.0018$), which is smaller than the observed depletion
\citep{2005AAS...20711411M}.

We have not further studied the effect of the surface abundances as a
function of the parameters. The trends of enhanced nitrogen and
depleted oxygen is promising, and we think that the relatively small
deviations from the observations can be matched by the simulation with
a more exhaustive exploration of the initial parameter space.  We
performed only a very limited parameter study, and a more exhaustive
study may be used to reduce the uncertainty in our adopted initial
conditions.

\subsection{The morphology of the outflow produced by the merger}

Apart from the ``Homunculus'' nebula the entire region around Eta
Carinae has a rather unstructured appearance \citep[see
  also][]{2007ApJ...666..967S}. Although our model of a binary merger
while orbited by a third companion tends to drive an asymmetric
outflow, at least as a result of the binary merger itself, there is
also a considerable component in which the outflow is symmetric. In
particular in the episode of highly enhanced wind loss prior to the
merger, as was discussed in sections \ref{Sect:TidalLuminosity} and
\ref{Sect:Brightening}, we expect to have produced a symmetric
bi-modal structure. We therefore think that the observed symmetries in
the Homunculus do not pose a serious limitation for our model (any
post-merger disturbance of the primary, e.g. by periastron passage of
the companion, will again lead to a Homonculus-like structure, such as
the Little Homonculus).

It should be noticed here that the merger product will rotate with
break-up velocity. Due to the Von Zeipel theorem
\citep{1924MNRAS..84..665V} the polar regions of this star will
therefore be very much hotter than its equatorial regions, which is
expected to lead to very much stronger radiatively driven stellar wind
mass loss from the polar regions than from the equatorial regions. We
suggest that this has produced the Homunculus structure of Eta Carinae
\citep[see also][]{1999A&A...347..185M,2004A&A...418..639A}.

From interferometric measurements \cite{2012ASSL..384..129W} noticed
that, apart from the bipolar-shaped Homunculus there are slow-moving
``knots'' --- condensations within 0.3'' from the central star that
are moving at $v \simeq 50$ km/s, and apparently have been ejected in
or soon after the great eruption, and seem ejected more or less
perpendicular to the axis of the bipolar nebula: in the equatorial
plane of the Homunculus.

This is confirmed by Weis' (2012)\nocite{2012ASSL..384..171W} study of
the HST and CHANDRA images of the surroundings of the star.  Weis
notices that, while the Homunculus with its two lobes (and equatorial
``skirt'' of $\aplt 1~\MSun$) has a near-perfect bipolar symmetry, the
outer ejecta that presumably date from the pre-1838 tidally induced
dense stellar wind, appears much more asymmetric, and is composed of
many irregularly-shaped structures in a roughly elliptically-shaped
region around the Homunculus, the axis of the ellipse being
perpendicular to the axis of the latter: the largest are the
S-condensation and the S-ridge (a large highly asymmetrically-shaped
structures on one side of the Homunculus). This condensation as well
as the NN-knots on the opposite side of the Homunculus were apparently
ejected perpendicular to the axis of the Homunculus (in its equatorial
plane). The fact that these ejecta are nitrogen enriched indicate that
CNO-processed material was brought to the surface and was violently
ejected in the great eruption.  Such nitrogen enrichment is a natural
consequence of the proposed merger scenario (see also
\S\,\ref{Sect:Discussion.sabundancy}), in which we presented the
result of one of our SPH simulation of the merger, and where we
indicate the presence of nitrogen enhanced material.
\cite{2012ASSL..384..171W} notices that these outer ejecta are more or
less a conglomerate of individual smaller structures and not a
coherent circumstellar shell: {\em``Its probable origin in the early
  nineteenth century .... makes formation of the outer ejecta by
  fragmentation of an expanding shell seem unlikely. Its morphology,
  high velocities and large velocity dispersion of individual
  structures, and the X-ray emission formed through shocks, support
  its creation in a more explosive event, during which several outer
  layers of the star's surface were ejected.''}

It appears from this that apart from the formation of the Homunculus
there has been a period of large asymmetric mass ejections in the axis
of the Homunculus, involving outer layers of the
star. \cite{2012ASSL..384..171W} estimates the combined mass of these
ejecta to be at least 2 -- 4\,\MSun, but such estimates are highly
uncertain and easily could be as large as 10\,\MSun\, or more.  We
present the amount of mass lost in the merger events in
Tab.\,\ref{Tab:HydroModels}. The mass of the inner nitrogen enriched
shell (the inner loop in fig.\,\ref{Fig:SPH_xy_post_merger}) comprises
about $\sim 2.9$\,\MSun, of the total ejected $\sim 22.2$\,\MSun.  In
the other hydrodynamical simulations a similar fraction of the ejected
mass (of about 15\%---25\%) is deposited in the nitrogen enriched
skirt.

The large asymmetric mass ejection from the equatorial regions of the
star (assuming the orbital plane of the original inner binary to
coincide with the equatorial plane) is precisely what our merger model
predicts. As a consequence, our model predicts that the current
orbital plane is alinged with the equatorial plane of the Homunculus,
and the symmetric lobes are roughtly aligned with the argument of
periastron of the current Eta Carina binary.

\subsection{Composition of the outflow}

In Fig.\,\ref{Fig:3DSPH_xy_post_merger} we notice that the inner loop
(blue in the figure) in spatial vs velocity coordinates is enhanced in
nitrogen by about a factor of 10 compared to the outer loop (red in
fig.\,\ref{Fig:3DSPH_xy_post_merger}). The same is the case for
carbon, but there the outer ring is composed of carbon poor material
(with an abundance of $<0.0002$) and the inner structure is enhanced
in carbon by more than an order of magnitude ($>0.003$).  A similar,
but less extreme, bimodal distribution is observed in the oxygen
content, but the range in abundance is much smaller.

The abundance differences across the remnant are due to the
inspiraling process of the two stars. In the first violent passage
mostly the outer layers are ejected, and upon each subsequent burst
mass from the deeper layers of the primary star are ejected.  Those
deeper layers are more enhanced in CNO processed material.

Shortly after the merger, the star is bloated to about 120\,\RSun, and
rapidly rotating (see \S\,\ref{Sect:EtaCar1838_1843}). We did not
follow the subsequent evolution of the merger product, because the
Henyey stellar evolution code will be unable to follow the non-thermal
evolution of the star properly. The current secondary star, which once
was the outer tertiary star, will plunge through this envelope at its
next pericenter passage, some 5.5\,yr after the merger event.

\section*{Acknowledgments}

It is a pleasure to thank Adrian Hamers, Tjarda Boekholt, Alex Rimoldi
and Arnout van Gelderen for discussions, and the anonymous referee for
grittycally reading the manuscript.  This work was supported by the
Netherlands Research Council NWO (grants \#643.200.503, \#639.073.803
and \#614.061.608) by the Netherlands Research School for Astronomy
(NOVA) and by the National Science Foundation under Grant No. NSF
PHY11-25915 Part of the numerical computations were carried out on the
Little Green Machine at Leiden University.

\begin{table}
  \caption{Reconstructed parameters for Eta Carinae.  The first column
    identifies the parameter, and the subsequent columns give
    estimates for the conditions at birth, during the 1838 merger
    event and today. The latter are taken from the literature
    \citep{2013MNRAS.429.2366S}.  The parameters are the age of the
    primary star at the moment of the collision, the total mass of the
    inner binary, the mass lost in that phase of the evolution, and
    the mass of the outer component (later to become the secondary
    star). We further give the inner binary orbital parameters
    ($a_{\rm in}$ and $e_{\rm in}$) and the outer elements ($a_{\rm
      out}$ and $e_{\rm out}$).  The last four parameters give the
    relative inclination of the two orbits, the mean anomaly of the
    outer orbit at the moment the merger, the kick velocity and its
    direction in the inner orbital plane.  }
  \label{tab:EtaCar}
  \begin{center}
  \begin{tabular}{llll}
  \hline
            & Phase A   & Phase B     & Phase C \\
  \hline
  parameter & pre 1838  & 1838        & post 1838 \\
  \hline
 $t$                 &                & 0.1---1.0\,Myr     &  $\aplt 1$\,Myr \\
 $M_{\rm in}+m_{\rm in}$ &120---150\MSun  &$114\pm 12.2$\MSun  &  90\MSun   \\
 $dm_{\rm in}$         &22.4\,\MSun    &$23.7\pm12.2$\MSun   &        \\
 $m_{\rm out}$         &30\,\MSun       &30\MSun              &    30\MSun  \\
 $a_{\rm in}$          &1---2\,AU      &0.36---0.8\,AU       &        \\
 $e_{\rm in}$          & 0--0.8        & 0---0.96            &      \\
 $a_{\rm out}$         &20---100\,AU   & 20---100\,AU        & 15.46\,AU \\
 $e_{\rm out}$         &0.02---0.7     &0.02---0.7           & 0.9       \\
 $i$                 & ---           & $91^\circ \pm 21^\circ$ & \\
mean anomaly         & ---           &$0^\circ.061 \pm 0^\circ.035$ &\\
$v_{\rm kick}$         & ---           &$53.1 \pm 7.7$\,km/s &\\
$\theta$             & ---           &$180^\circ \pm 9^\circ$&\\
  \hline
  \end{tabular}
  \end{center}
\end{table}

\end{document}